\documentclass[prb, twocolumn, nofootinbib, superscriptaddress, citeautoscript, nobibnotes, nofootinbib]{revtex4-2}

\usepackage{comment}

\usepackage[]{standalone}
\usepackage{import}
\usepackage{booktabs}
\usepackage{tabularx}
\usepackage{array}

\newcolumntype{Y}{>{\centering\arraybackslash}X}
\usepackage[utf8]{inputenc}
\usepackage[T1]{fontenc}
\usepackage{quantikz}
\usepackage{cmap}
\usepackage{graphicx}
\usepackage{mathtools}

\usepackage[hyperfootnotes=false,breaklinks=true]{hyperref}
\usepackage{setspace}
\hypersetup{
 bookmarksopen=true,
 bookmarksopenlevel=1,
 colorlinks=true,
 linkcolor=blue,
 anchorcolor=blue,
 citecolor=blue,
 filecolor=blue,
 urlcolor=blue,
 pdfpagemode=UseOutlines,
 pdfstartview={XYZ null null 1},
 linktocpage=true,
}
\usepackage{amsmath}
\allowdisplaybreaks  
\usepackage{amssymb}
\usepackage{amstext}
\usepackage{amsfonts}
\usepackage{mathrsfs}
\usepackage{amsthm}
\usepackage{dsfont}
\usepackage{bm}
\usepackage{braket}
\usepackage{physics}
\usepackage{enumitem}
\usepackage[
 capitalise,
]{cleveref}

\usepackage{pgfplots}
\pgfplotsset{compat=newest}
\usepackage{tikz}
\usetikzlibrary{
  positioning,
  shapes.geometric,
  arrows,
  arrows.meta,
  backgrounds,
  fit,
  intersections,  
}
\usepackage[english]{babel}
\usepackage{algorithm,algorithmic}

\theoremstyle{plain}

\theoremstyle{definition}

\newcommand{\mycomment}[1]{}

\usepackage{relsize}
\usepackage{cancel}

\newcommand{\opbasis}{\mathcal{P}}

\newcommand{\PREP}{\texttt{PREP}}
\newcommand{\SELECT}{\texttt{SELECT}}
\newcommand{\PHASE}{\texttt{PHASE}}
\newcommand{\CPREP}{\texttt{CPREP}}

\makeatletter
\def\blfootnote{\xdef\@thefnmark{}\@footnotetext}

\begin{document}

\blfootnote{This manuscript has been authored by UT-Battelle, LLC, under Contract No. DE-AC0500OR22725 with the U.S. Department of Energy. The United States Government retains and the publisher, by accepting the article for publication, acknowledges that the United States Government retains a non-exclusive, paid-up, irrevocable, worldwide license to publish or reproduce the published form of this manuscript, or allow others to do so, for the United States Government purposes. The Department of Energy will provide public access to these results of federally sponsored research in accordance with the DOE Public Access Plan.}

\title{Matrix Product Operators In The Age of Block Encoding}

\author{Eugene Dumitrescu}  
\email{dumitrescuef@ornl.gov}
\affiliation{Computational Sciences and Engineering Division, 
Oak Ridge National Laboratory, 
Oak Ridge, Tennessee 37831, USA}

\date{\today}
\begin{abstract}

We develop a block-encoding compiler that treats matrix product operators as compressed, virtual-path linear combination of unitaries programs. The compiler constructs conditional \texttt{PREP} and local \texttt{SELECT} stages directly from a parent matrix product operator, establishing tensor networks as structured quantum intermediate representations that can be efficiently compiled to block-encoded circuits. We apply the construction to real-time evolution in the Heisenberg chain and two perturbed Heisenberg-family models. Across the regimes studied, the compressed, approximately unitary propagator MPOs retain mild bond dimension and LCU normalization. Relative to an LCU that explicitly materializes the \(\mathcal{O}(N^K)\) Pauli-product branches of an order-\(K\) truncated Taylor polynomial, our virtual-transition implementation replaces combinatorial branch enumeration by a circuit complexity scaling as \(\mathcal{O}(\alpha_{\rm MPO}N\chi^2)\), approaching \(\mathcal{O}(N\chi^2)\) when \(\alpha_{\rm MPO}\) remains mild. We numerically characterize how truncation order, bond-dimension budget, and system size affect approximation error, normalization, and compiler cost. These results demonstrate how classically compressed tensor-network representations can serve as quantum compiler intermediate representations for block encoding and opens new avenues to accelerate quantum algorithms.

\end{abstract}
\maketitle

\section{Introduction}
\label{sec:motivation}

Quantum representation, characterization, control, compilation, and computation are difficult, multifaceted tasks. Optimizing these tasks and reasoning about the large number of degrees of freedom comprising quantum systems is often limited by the lack of guidance from classical data structures. Recently, quantum simulation algorithms have increasingly exploited specialized classical data structures that expose the intrinsic algebraic, fermionic, tensorial, or low-rank structure of quantum models. These have led to exponential circuit-depth improvements~\cite{Wan2021}, polynomial reductions in oracle complexity~\cite{Babbush2018, Berry2019, Lee2021}, and large resource savings from compressed factorizations and optimized data-loading backends~\cite{Sanders2019, Oumarou2024, Rocca2024, Low2024Dirty, Sunderhauf2024, Caesura2025}.

Recently, tensor network methods have outgrown their traditional role as compressed multilinear data structures for quantum many-body states and operators. They now appear across a broad set of quantum protocols, including many-body tomography and characterization~\cite{Cramer2010,Guo2024,Qin2024}, non-Markovian process modeling~\cite{Strathearn2018,Pollock2018,Fux2023}, state preparation~\cite{Melnikov2023,Malz2024,Smith2024,Niall2025,wang2025}, optimal control~\cite{Doria2011,BaladaGaggioli2025}, predicting algorithmic performance under structured noise~\cite{Dumitrescu2017,Dang2019,jamadagni2026}, and distributed quantum simulations~\cite{Dwivedi2026DistributedTN}.

Tensor network methods have likewise entered quantum programming itself, with recent work introducing tensor quantum programming as a route to matrix-encoded quantum algorithms~\cite{Termanova2024}. However, much of tensor-network-based quantum compilation has treated tensor networks primarily as compact objects to be variationally converted into circuits, rather than as compiler data structures for assembling block-encoding oracles. This distinction matters because modern block-encoding algorithms require more than a compact matrix representation: they require normalization accounting, amplitude-amplification costs, PREPARE/SELECT structure, oracle composition rules, and compiler-level resource estimates. What is missing is compiler analysis that treats tensor networks themselves as a structured intermediate representation for quantum linear algebra.

Complementary work has shown that MPOs~\cite{Nibbi2024}, and more general tensor-network operators~\cite{Issel2026}, can be placed inside block-encoding, that tensor networks can serve as state-preparation circuits~\cite{Sopena2022}, and that tensor-network block encodings find use in differential-equation solvers~\cite{Siegl2026}. These results demonstrate that tensor networks can be mapped into quantum circuits or embedded into block-encoded quantum linear-algebra routines. However, they do not address the higher-level compiler question: how can tensor networks be used not merely as compact matrices to be block encoded, but as structured data from which block-encoding oracles can be systematically assembled, analyzed, and optimized? 

In this work we make three contributions. First, we show that an MPO can be interpreted as a compressed LCU over virtual paths. Second, we give a block-encoding compiler based on conditional preparation, phase application, and local selection oracles. Third, using Hamiltonian simulation as a case study, we compare noncommuting compilation orders and show that compiling the target polynomial at the MPO level avoids {\em explicit} Pauli-branch complexity and, for the MPO gauges studied here, the normalization growth observed for the $H$-first local tensor-dilation routes~\cite{Nibbi2024}. We conclude with a discussion of the future directions our work opens. 

\section{Overview of Main Results}
\label{sec:MainResults}

The main idea behind our MPO compilation is as follows. Instead of directly compiling to an explicit sum of Pauli operators, we seek to speed up block encodings by exploiting correlations among the local Pauli choices that combine into long Pauli strings. The MPO is a data structure that naturally compresses such correlations and can be used to block encode various operators with costs polynomial in the bond dimension~\cite{Nibbi2024, Issel2026, Siegl2026}. However, even within the realm of MPO block encoding, there are several inequivalent compilation routes. Existing MPO block-encoding analyses often treat the MPO as an object to be locally unitarized.

That is, a tensor with virtual bond dimension ($\chi$) is reshaped into a matrix acting on physical and virtual-bond registers, and is then embedded into a larger unitary dilation. This gives a valid block encoding, but it also commits the compiler to the normalization and circuit costs induced by that local dilation. In practical tensor-network compilation, by contrast, the MPO may itself serve as an approximate, compressed, and optimizable representation of the target transformation. Our method avoids committing to a local unitary dilation of the Hamiltonian MPO. Instead, we compile the target polynomial transformation within the MPO representation before performing the final block encoding. \[ \begin{array}{ccc} 
H_{\rm MPO} & \xrightarrow{\text{compile }p(H)\text{ as MPO}} & p(H)_{\rm MPO} \\
[4pt] \downarrow{\text{BE}} && \downarrow{\text{BE}} \\[4pt] {\rm BE}(H) & \xrightarrow{\text{QET}~\cite{Nibbi2024}} & {\rm BE}(p(H)) 
\end{array} \] In this work, we highlight compilation choices of this kind and show that they can dramatically affect performance and resource costs.

The relevant comparison is therefore between algorithms that target the same transformation, but are inequivalent in their resource requirements,
\begin{equation}
\label{eq:pauli_expansion}
P_K \approx e^{-itH} \xrightarrow{\rm Pauli \;  Expansion} {\rm BE}(P_K(t)),
\end{equation}
\begin{equation}
\label{eq:Nibbi_route}
    H_{\rm MPO}\to {\rm BE}(H)\to {\rm QET}(e^{-itH}),
\end{equation}
and 
\begin{equation}
\label{eq:MPO_expansion}
   H_{\rm MPO}\to P_\chi(t)\approx e^{-itH}\to {\rm BE}(P_\chi(t)).
\end{equation}
The first route, given by Eq.~\ref{eq:pauli_expansion}, denotes a flat compilation in which the order-$K$ polynomial is materialized as an explicit list of Pauli-product branches. For a Hamiltonian $H=\sum_{\ell=1}^{T}h_\ell P_\ell$, the uncollected order-$k$ contribution contains $T^k$ Pauli-product histories, and the corresponding flat branch table contains up to $\sum_{k=0}^{K}T^k=\mathcal{O}(T^K)$ entries. This explicit Pauli enumerated route is the Pauli-LCU baseline used in our scaling comparisons.

This baseline should not be identified with all Taylor-series LCU algorithms. For example, the BCCKS construction~\cite{BCCKS2017} exploits the special Cartesian-product factorization of the Taylor-history coefficients, $\beta_{k,\ell_1,\ldots,\ell_k} = \frac{(t/r)^k}{k!} \prod_{j=1}^{k}|h_{\ell_j}|$. Its \texttt{PREP} operation therefore factorizes into an order-register preparation and $K$ term-index preparations, while its \texttt{SELECT} operation factorizes into $K$ successive controlled-(\texttt{SELECT}(H)) stages. BCCKS consequently avoids explicitly materializing or independently processing the $T^K$ Taylor histories. We do not claim an improvement over this structured construction. Instead, Eq.~\ref{eq:pauli_expansion} isolates the compiler benefit of retaining the final polynomial operator as a compressed MPO rather than flattening it into an explicit Pauli-product enumerated table. The two approaches exploit different structures. BCCKS factorizes the algorithmic Taylor histories, whereas the present compiler, outlined below, factorizes the spatial operator coefficients through MPO virtual bonds. 

The second route, given by Eq.~\ref{eq:Nibbi_route}, first block encodes $H$ as an MPO~\cite{Nibbi2024} with a dilation $U_H$ such that $(\langle 0|\otimes I)U_H(|0\rangle\otimes I) =  \frac{H}{\alpha_{\rm dil}(H)}$. QET then approximates \(e^{-it\alpha_{\rm dil}(H) x}\), so the polynomial degree is controlled by \(D(\alpha_{\rm dil}(H) t,\epsilon)\). Hence, the first route, in Eq.~\ref{eq:Nibbi_route}, scales as the cost to block encode $H$ and to then build polynomials querying the $H$ block encoding $C_{\ref{eq:Nibbi_route}} \sim D(\alpha_{\rm dil}(H) t,\epsilon)\, C_{\rm BE}(H).$ That is, the unitary-dilated Hamiltonian block-encoding normalization \(\alpha_{\rm dil}(H)\) appears inside the QET degree \(D(\alpha_{\rm dil}(H) t,\epsilon)\). In the normalized Heisenberg MPO gauge, we compute a Ref.~\citenum{Nibbi2024}-style normalization proxy by taking the product of the local spectral norms of the reshaped MPO tensors. For our gauge, we find $ \alpha_{\rm dil}(H)^{\rm proxy} \approx 0.232(1.43)^N.$ This exponential fit comes from our implementation of their local tensor-dilation normalization on the MPO family studied here and is consistent with the prior published analysis~\cite{Nibbi2024}. Thus, the query-polynomial route inherits an exponential normalization penalty.

The third route, given by Eq.~\ref{eq:MPO_expansion}, pays the block-encoding normalization of the compiled approximately unitary MPO, \(\alpha_{\rm MPO}(P_\chi)\). Our numerical data indicates (Sec.~\ref{sec:apps}) that \(\alpha_{\rm MPO}(P_\chi)\) remains mild for the normalized Heisenberg Taylor MPO expansions of real-time evolution unitaries. The total cost is $C_{\ref{eq:MPO_expansion}} \sim \alpha_{\rm MPO}(P_\chi(t))\, C_{\rm BE}(P_\chi(t))$. The conclusion is that this direct route avoids the exponential normalization bottleneck associated with the MPO block encoding of non-unitary $H$~\cite{Nibbi2024}. Here, the compiled-MPO route first constructs a compressed approximately unitary MPO $ P_\chi(t)\approx e^{-itH}$ and then block encodes \(P_\chi(t)\). Across the tested \((N,\epsilon)\) grid, the MPO-transition normalization \(\alpha_{\rm MPO}(P_\chi)\) remains bounded by a small constant factor, and the circuit proxy is well described by $C_{\ref{eq:MPO_expansion}} \sim \alpha_{\rm MPO}(P_\chi)N\chi_P^2.$ Therefore, for the tested regimes, compiling the unitary-valued polynomial at the MPO level simultaneously avoids the exponential normalization bottleneck of the $H$ block encoding route and the combinatorial bottleneck of the baseline explicit Pauli-enumerated LCU expansions.

\section{Block Encoding}
\label{sec:block_intro}
In this work we compare three distinct block-encoding routes: i) explicit Pauli LCUs, ii) tensor-dilation~\cite{Nibbi2024} in which local MPO tensors are unitarized, and iii) MPO-transition LCUs where MPOs are expanded into local Pauli transitions. Linear combination of unitaries (LCU) is a family of quantum algorithms that apply an operator 
\begin{equation}
    \label{eq:LCU}
    A=\sum_j c_j U_j,
\end{equation}
to a quantum state. The block-encoding normalization is $\alpha = \sum_j |c_j|.$ A standard example that comes from expanding a Hamiltonian in a (unitary) Pauli basis, 
\begin{equation}
    \label{eq:pauli_ham}
    H=\sum_{\ell=1}^{T} h_\ell P_\ell.
\end{equation}
For the normalized generator \(\bar H\), we use
\begin{equation}
\label{eq:polynomial}
p_K(\bar H)=\sum_{k=0}^{K} c_k \bar H^k    
\end{equation}
an explicit Pauli expansion can grow combinatorially. An important example is the Taylor expansion of the real time evolution $e^{-i\bar Ht}$, with $c_k = \frac{(-i t)^k}{k!}$,  or imaginary $e^{-H\tau}$ time evolutions~\cite{BCCKS2017}. In such cases, a explicitly enumerated Pauli LCU representation of polynomials $p_K(\bar H)$ has a Pauli count $\sum_{k=0}^{K}T^k = O(T^K)$. For a local spin chain, with extensive $T=O(N)$ interactions, this gives $O(N^K)$ Pauli complexity to the compilation route given in Eq.~\ref{eq:pauli_expansion}.

\subsection{Normalization and Success Probability}
\label{sec:norm_and_success}

For a geometrically local Hamiltonian \(H_N\), the natural Pauli-LCU normalization is 
\begin{equation}
    \label{eq:norm}
    \alpha_{\rm P}(H_N) = \sum_{\ell}|h_\ell| = \mathcal{O}(N).
\end{equation}
Using this, define the normalized Hamiltonian $\bar H_N=\frac{H_N}{\alpha_{\rm P}(H_N) }.$ One then utilizes the normalized generator \(\bar H_N\) and synthesize $U(t)=e^{-it\bar H_N}.$ Thus, a specified \(t\) corresponds to fixed normalized time. Physical evolution \(e^{-iT H_N}\) corresponds to \(t=T\alpha_{\rm P}(H_N) \). Therefore the  fixed-\(t\) scaling statements should be interpreted as compiler scaling at fixed normalized time; long-time or fixed-physical-time scaling is a separate question controlled by the \(t\)-dependence of the compiled MPO. This Hamiltonian normalization should be distinguished from the normalization of the full explicit Pauli expansion of a polynomial transformation. If \(P_K(t)=\sum_{s\in\mathcal S_K} d_s P_s \approx e^{-it\bar H_N}\), then the explicit Pauli-LCU route pays the polynomial normalization \(\alpha_{\rm P}(P_K)=\sum_{s\in\mathcal S_K}|d_s|\). It is this latter quantity, together with the cost of preparing and selecting from \(\mathcal S_K\), that enters the \(C_{\ref{eq:pauli_expansion}}\) comparison. The role of Eq.~\ref{eq:norm} is instead to define the normalized generator and the corresponding time scale.

Before proceeding into the technical details, let us briefly comment on the algorithmic impact of $\alpha$ on the target operator. If $A$ is approximately unitary, then $\|A|\psi\rangle\|\approx 1,$ and the success probability is approximately $ p_{\mathrm{succ}}\approx \frac{1}{\alpha^2}$. Thus $\alpha$ is a representation overhead. In general, if $A$ is nonunitary, then $p_{\mathrm{succ}} = \frac{\|A|\psi\rangle\|^2}{\alpha^2}$. The cost depends both on the representation normalization $\alpha$ and on the intrinsic norm change of the operator on the input state. This distinction is important for bare Hamiltonian application, imaginary-time evolution, filters, projectors, and dissipative maps. Thus, repeated post-selection costs \(O(\alpha^2)\), while amplitude amplification costs \(O(\alpha)\)~\cite{Brassard2000}. In the cost proxies below we use the amplitude-amplified scaling, so \(\alpha\) multiplies the circuit-call cost. For nonunitary operators, \(\alpha\) is not the whole story because \(\|A|\psi\rangle\|\) may itself be small. We do not address this case in this work and only briefly comment on it in the conclusion. 

\section{Virtual Path LCU}
Tensor networks are tools to compress a variety of low-rank operators~\cite{Orus2019}. Before directly translating the MPO data structure into a block-encoded quantum circuit, we briefly review this data structure and quantum type's formalism. 

\subsection{Notation}
Let $M$ be an MPO acting on $N$ sites. The MPO is a linear operator taking physical input indices $t_n$ to physical output indices $s_n$. We write
\begin{equation}
\label{eq:MPO}
    M = \sum_{\mathbf{s},\mathbf{t}} \sum_{\mathbf{a}} M^{[1]\,s_1,t_1}_{a_1} M^{[2]\,s_2,t_2}_{a_1,a_2} \cdots M^{[N]\,s_N,t_N}_{a_{N-1}}  |\mathbf{s}\rangle\langle\mathbf{t}|.
\end{equation}
The physical degrees of freedom factor locally. At site \(n\), define the operator \( O^{[n]}_{a,b} = \sum_{s,t}M^{[n]\,s,t}_{a,b}|s\rangle\langle t|. \) The MPO is obtained by contracting the virtual indices: \( M = \sum_{a_1,\ldots,a_{N-1}} O^{[1]}_{a_1} O^{[2]}_{a_1,a_2} \cdots O^{[N]}_{a_{N-1}}. \)

\subsection{Local unitary-basis expansion}
Unitary operators are useful in constructing block encodings~\cite{Childs2012}. Choose a local unitary operator basis $ \opbasis=\{P_\mu\}_{\mu=1}^{q}$ which, for qubits, reduces to the Pauli basis $\opbasis=\{I,X,Y,Z\}$ with $q=4$. For qudits, $q=d^2$. Expand each local operator block as $O^{[n]}_{a,b} = \sum_{\mu=1}^{q} W^{[n],\mu}_{a,b} P_\mu^{(n)}.$ For local dimension $d$,
\begin{equation}
\label{eq:pauli_transfer_matrices}
W^{[n],\mu}_{a,b}  = \frac{1}{d} \operatorname{Tr} \left[ P_\mu^\dagger O^{[n]}_{a,b} \right].    
\end{equation}
which is a transfer matrix for $P_\mu$ at site $n$.

\subsection{Virtual paths}

One central observation of this work is that an MPO can be interpreted directly as a compressed LCU program. Rather than treating each MPO tensor as a dense local map to be unitarized~\cite{Nibbi2024}, we instead expand each local tensor in a local unitary operator basis. In this construction we interpret the virtual bond as a finite-state transition automaton~\cite{Crosswhite2008}. This gives an LCU over virtual paths rather than over explicitly materialized Pauli strings. Below, we outline this construction. 

Modulo global $\pm 1, \pm i$ phases, the $N$-qubit Pauli group is $\mathbb{P}_N/\{\pm 1, \pm i\} = \opbasis^{\otimes N}$. Let us examine how specific Paulis, say the product $ P_\gamma = P_{\mu_1}^{(1)} P_{\mu_2}^{(2)} \cdots P_{\mu_N}^{(N)}$, are encoded within the MPO-mediated block encodings. First, define a virtual-transition path,
\begin{equation}
\label{eq:VTPath}
    \gamma=(\mu_1,a_1,\mu_2,a_2,\ldots,a_{N-1}, \mu_N).   
\end{equation}
For each virtual bond space, \(V_n\), choose a basis $\bigl\{|a_n\rangle\bigr\}$ with $ \bigl\{ a_n=1,\ldots,\chi_n\bigr\}.$ Here, \(a_n\) denotes a basis label, not the entire virtual space \(V_n\). Thus, a virtual path is a \textit{particular sequence} of basis labels together with local operator labels. The coefficient associated with the (open boundary condition) $\gamma$ path is
\begin{equation}
\label{eq:path_coeff}
c_\gamma =  W^{[1],\mu_1}_{a_1} \left( \prod_{n=2}^{N-1} W^{[n],\mu_n}_{a_{n-1},a_n} \right) W^{[N],\mu_N}_{a_{N-1}}.
\end{equation}

The MPO in Eq.~\ref{eq:MPO} is therefore re-expressed as 
\begin{equation}
\label{eq:MPO_Pauli_expansion}
    M = \sum_{\gamma\in\Gamma} c_\gamma P_\gamma
\end{equation}
where $\Gamma$ denotes the set of all paths and the virtual bond has become a finite-state automaton generating Pauli strings. 

\subsection{Normalization}

Before proceeding to explicit constructions, let us examine normalization in the context of MPO-based LCU. If the block encoding is of $T/\alpha$ and $\alpha$ is huge, success probability or QET scaling may be bad. For MPOs, local tensor normalization impacts the algorithmic performance. Nibbi et al.~\cite{Nibbi2024} explicitly point out that their method needs per-tensor normalization, which rescales the Hamiltonian.

From Eq.~\ref{eq:MPO_Pauli_expansion} we immediately see the coefficient norm of the tensor-network branch decomposition is
\begin{equation}
    \alpha_{\rm MPO} = \sum_{\gamma\in\Gamma}|c_\gamma|
\end{equation}    
where $\Gamma$ denotes the set of all paths. 

For a fixed path \(\gamma\), all virtual labels \(a_n\) and local operator labels \(\mu_n\) have been chosen. Therefore each factor $W^{[n],\mu_n}_{a_{n-1},a_n}$ is a complex scalar matrix element, not a matrix. Projecting  Eq.~\ref{eq:pauli_transfer_matrices} into a particular path reveals a complex coefficient such that,  \( |c_\gamma| = \left| \prod_{n=1}^{N} W^{[n],\mu_n}_{a_{n-1},a_n} \right| = \prod_{n=1}^{N} \left| W^{[n],\mu_n}_{a_{n-1},a_n} \right|. \) This equality is exact because the $W$  factors are complex numbers. It would not be valid if \(W^{[n],\mu_n}\) denoted the full virtual-space matrix and the absolute value were interpreted after matrix multiplication or virtual-index contraction.

For \(2\le n\le N-1\), define the absolute transition matrices \( A^{[n]}_{a_{n-1},a_n} = \sum_{\mu_n}|W^{[n],\mu_n}_{a_{n-1},a_n}|. \)  with \( A^{[1]}_{a_1} = \sum_{\mu_1}|W^{[1],\mu_1}_{a_1}|, \qquad A^{[N]}_{a_{N-1}} = \sum_{\mu_N}|W^{[N],\mu_N}_{a_{N-1}}|, \) at the boundaries. Then 
\begin{equation}
\label{eq:transition_matrix_alpha}
\alpha_{\rm MPO} = \sum_{a_1,\ldots,a_{N-1}} A^{[1]}_{a_1} \left( \prod_{n=2}^{N-1} A^{[n]}_{a_{n-1},a_n} \right) A^{[N]}_{a_{N-1}}
\end{equation}

Equation~\ref{eq:transition_matrix_alpha} gives the LCU normalization induced by the MPO representation. Furthermore, importantly, the normalization is efficiently computed by tensor contraction which does not require one to sum over all paths independently. This norm should be distinguished from the coefficient norm of the explicitly collected Pauli decomposition \( \alpha_{\rm P}(P_K)\) discussed below Eq.~\ref{eq:norm}. The two decompositions represent the same operator, but their branching and compilation strategies are different. If several MPO paths contribute to the same Pauli string, then \( d_s = \sum_{\gamma:P_\gamma=P_s}c_\gamma. \) Thus \( \alpha_{\rm MPO} = \sum_{\gamma}|c_\gamma| \ge \sum_s \left| \sum_{\gamma:P_\gamma=P_s}c_\gamma \right| = \alpha_{\rm P}(M)\). In the numerical examples of Sec.~\ref{sec:apps}, this normalization overhead remains modest and leads to an overall reduction in circuit complexity relative to the fully enumerated Pauli baseline.

\subsection{High-Level Compilation}

The typical LCU construction is $U_{\mathrm{LCU}} = \PREP^\dagger \SELECT \,\PREP.$ 
\begin{quantikz}[row sep=0.45cm, column sep=0.55cm]
\lstick{$\ket{0}$}
& \gate{\PREP}
& \gate[wires=2]\SELECT
& \gate{\PREP^\dagger}
& \meter{} \\
\lstick{$\ket{\psi}$}
& \qw
&
& \qw
& \qw
\end{quantikz}
Here \(\PREP\) prepares amplitudes proportional to \(\sqrt{|c_\ell|}\) and \(\SELECT\) applies the phase and unitary associated with branch \(\ell\).
Targeting the application of Eq.~\ref{eq:LCU}, $\PREP|0\rangle = \sum_{\ell=1}^{T} \sqrt{\frac{|c_\ell|}{\alpha}} |\ell\rangle $ and $\SELECT = \sum_{\ell=1}^{T} |\ell\rangle\langle \ell| \otimes e^{i\arg(c_\ell)}U_\ell$.

This structure will be maintained, due to commutivity structure explained below, but the components of $U_{\rm LCU}$ will be generalized to construct the MPO paths unitary. Namely, $\PREP = \texttt{COND-PREP}_N \cdots \texttt{COND-PREP}_2 \texttt{PREP}_1$ factorizes into a series of $N$ conditional preparations. The select circuit likewise factorizes, $\SELECT = \SELECT_1 \SELECT_2 \cdots \SELECT_N.$ As it only depends on the \PREP of the same stage $\SELECT_i$ may be applied immediately after $\CPREP_i$. Using the key fact that it commutes with later $\CPREP_m$ operators, $\SELECT_n$ can however be applied immediately after $\CPREP_n$. But note that $\CPREP_i^\dagger$ cannot be applied immediately after $\SELECT_i$, because future steps (e.g. $\SELECT_{i+j}$) need the virtual path, which is established by later \PREP stages. Altogether, using the $\leftarrow$ symbol to denote that indices are time-ordered with later values coming to the left, we have 
\begin{align}
    U & = (\prod_{\leftarrow j} \CPREP_j)^\dagger  (\prod_{\leftarrow i} \SELECT_i \CPREP_i) \\
    & = (\prod_{\leftarrow j} \CPREP_j)^\dagger  (\prod_{\leftarrow k} \SELECT_{k})  (\prod_{\leftarrow i} \CPREP_i) \nonumber \\ 
    &= \PREP^\dagger \SELECT \,\PREP \nonumber
\end{align}

The reason for this generalization over the usual circuits is because, to block encode the MPO, we must sum over all paths by with a quantum circuit that effectively contracts the transfer matrices. At each site \CPREP is multiplexed state preparation indexed by incoming virtual label $a$. Using generic dense multiplexed state preparation and counting arbitrary logical rotations as elementary operations, the baseline preparation cost is $\mathcal{O}(q\chi^2)$ per bulk site. This is a dense logical synthesis bound rather than a fault-tolerant gate count. Sparsity, low-rank transition structure, or further tensor factorization should be investigated in a later work.

Let us examine one $\SELECT_n \CPREP_n$ pair for a bulk site. First recall that our LCU is utilizes an ancilla bank that applies operators as conditioned on paths. Thus the ancillary system is initially $\ket{0}_{\Gamma}$ and is to be prepared with the appropriate coefficients in the $\ket{\mu_1,a_1,\mu_2,a_2,\ldots,a_{N-1}, \mu_N }$ basis. The conditional preparation has the form \[ \CPREP_n = \sum_a |a\rangle\langle a|_{a_{n-1}} \otimes \PREP_{n,a}^{(\mu_n,a_n)} ,\] where \(\PREP_{n,a}^{(\mu_n,a_n)}\) prepares the outgoing label \((\mu_n,a_n)\) conditioned on the incoming virtual label \(a=a_{n-1}\). A single local stage may be represented schematically as follows.
\[
\begin{quantikz}[row sep=0.35cm, column sep=0.45cm]
\lstick{$\ket{a_{n-1}}$}   & \gate[wires=3]{\texttt{COND-PREP}_n} & \gate[wires=3]{\texttt{PHASE}_n} & \qw        & \qw \\
\lstick{$\ket{0}_{\mu_n}$} &                                        &                                    & \ctrl{2}   & \qw \\
\lstick{$\ket{0}_{a_n}$}   &                                        &                                    & \qw        & \qw \\
\lstick{$\ket{\psi_n}$}    & \qw                                   & \qw                               & \gate{P_{\mu_n}^{(n)}} & \qw
\end{quantikz}
\]

\subsection{Encoding Conditional Probabilities}
\label{sec:enc_cond_prob}

Having fine-grained, from the global circuit to the components comprising the MPO-based LCU circuit, let us now examine the preparation stage in even greater detail. The goal of this circuitry is to first encode tensor-network branches, with probabilities proportional to the absolute value of their LCU coefficients, and to subsequently apply the corresponding controlled unitaries to the target register.

For an open-boundary MPO, a branch $\gamma=(\boldsymbol{\mu},\mathbf{a})=(\mu_1,\ldots,\mu_N;\,a_1,\ldots,a_{N-1})$ comes with a coefficient $c_\gamma = W^{[1],\mu_1}_{a_1} \left( \prod_{n=2}^{N-1} W^{[n],\mu_n}_{a_{n-1},a_n} \right) W^{[N],\mu_N}_{a_{N-1}}$ and the global MPO-transition LCU coefficient norm is $ \alpha_{\rm MPO} = \sum_{\gamma\in\Gamma_{\rm TN}} |c_\gamma|.$ To prepare branches, according to $ p(\gamma)=\frac{|c_\gamma|}{\alpha_{\rm MPO}},$ we inductively define backward messages $\beta_i$. At the last site, $ \beta_N(a_{N-1}) = \sum_{\mu_N} \left| W^{[N],\mu_N}_{a_{N-1}} \right|.$ For bulk sites \(2\le n\le N-1\), $ \beta_n(a_{n-1}) = \sum_{a_n,\mu_n} \left| W^{[n],\mu_n}_{a_{n-1},a_n} \right| \beta_{n+1}(a_n).$ Then, $\alpha_{\rm MPO} = \sum_{a_1,\mu_1} \left| W^{[1],\mu_1}_{a_1} \right| \beta_2(a_1).$

The backward messages \(\beta_n\) play the role of partial norms in a recursive state-preparation procedure. The total coefficient norm \(\alpha_{\rm MPO}\) normalizes the first preparation, while each subsequent \(\CPREP_n\) divides by the remaining suffix weight \(\beta_n(a_{n-1})\). Thus the product of conditional probabilities telescopes to $ p_{\CPREP}(\gamma)=\frac{|c_\gamma|}{\alpha_{\rm MPO}}.$ In this sense, the MPO contraction supplies the partial norms needed to factor a global branch distribution into local conditional preparations.

The first preparation is $$\CPREP_1: |0\rangle = \sum_{\mu_1,a_1} \sqrt{ \frac{ |W^{[1],\mu_1}_{a_1}|\beta_2(a_1)}{\alpha_{\rm MPO}}}|\mu_1,a_1\rangle.$$ For \(2\le n\le N-1\), the bulk conditional preparation is
\begin{align}
\CPREP_n  |a_{n-1} |0\rangle  =  |a_{n-1}\rangle \sum_{\mu_n,a_n} \Omega_{\mu_n,a_n} |\mu_n,a_n\rangle.
\end{align}
where $\Omega_{\mu_n,a_n} = \sqrt{ \frac{ |W^{[n],\mu_n}_{a_{n-1},a_n}| \beta_{n+1}(a_n)}{\beta_n(a_{n-1})}}.$ Finally, at the last site, $$ \CPREP_N |a_{N-1}\rangle |0\rangle = |a_{N-1}\rangle \sum_{\mu_N} \sqrt{ \frac{ |W^{[N],\mu_N}_{a_{N-1}}| }{ \beta_N(a_{N-1}) } } |\mu_N\rangle.$$

\section{Applications and Scaling Analysis}
\label{sec:apps}
In this section we provide concrete data for specific examples. We will focus on compiling the Heisenberg model and related non-integrable variants.

\subsection{Hamiltonian Simulation}
\label{sec:RTE}

For our central example, we consider the paradigmatic and universal task of Hamiltonian simulation. Specifically, we will compile the time evolution for Heisenberg-adjacent models in term of a power series expansion as suggested in Sec.~\ref{sec:block_intro}. Using $S^\mu_i=\sigma^\mu_i/2$ and $\bm{S}_i = (S^x_i, S^y_i, S^z_i)$ to denote the spin-1/2 operators, consider the Heisenberg model,
\begin{equation}
\label{eq:Heisenberg}
    H (J,N) = J \sum_{i=1}^{N-1} \bm{S}_i\cdot \bm{S}_{i+1}. 
\end{equation}

Our goal is now to synthesize the real time evolution unitary that is generated by these Hamiltonians. Using Eq.~\ref{eq:polynomial}, this is straightforwardly accomplished as a Taylor expansion up to an order $K$. Naively, for the explicit collected Pauli LCU baseline, write $e^{-i\bar H t} \approx p_K(\bar H) = \sum_{s\in \mathcal S_K} d_s P_s$ with $ \alpha_{\rm P}(p_K) = \sum_{s\in\mathcal S_K}|d_s|.$ This leads to an explicit Pauli LCU cost $C_{\ref{eq:pauli_expansion}} = \alpha_{\rm P}(p_K)\,|\mathcal S_K|.$ Equivalently, if one includes a fixed per-string SELECT cost, this becomes $C_{\ref{eq:pauli_expansion}} \sim \alpha_{\rm P}(p_K)\, C_{\rm SELECT}^{\rm Pauli}(|\mathcal S_K|).$ For the scaling comparisons below, we use $|\mathcal S_K|=O(T^K)=O(N^K)$, so that, for fixed normalized time and fixed \(K\), $C_{\ref{eq:pauli_expansion}}=O(N^K)$ up to the scalar Taylor normalization and model-dependent collection constants.

\begin{figure}
    \centering
    \includegraphics[width=1\linewidth]{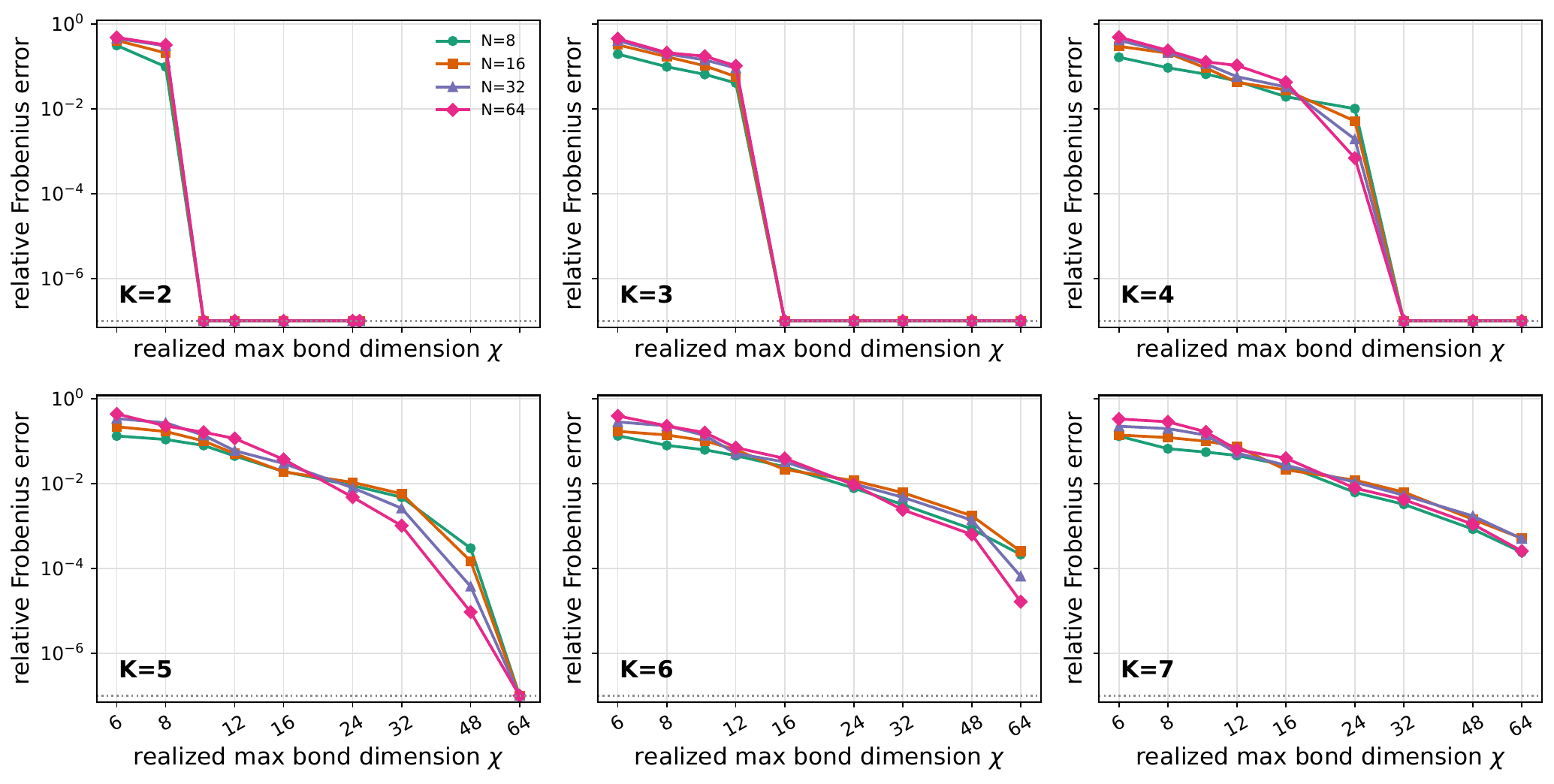}
    \caption{MPO bond dimension scaling for the representation of Hamiltonian powers. The bond dimension exhibits modest growth in $\chi$, significantly below the uncompressed $5^K$ scaling, and is importantly independent of system size. }
    \label{fig:HamiltonianCompression}
\end{figure}

However, our MPO construction differs. We first construct $H$ as an MPO that is compressed. We find that the Heisenberg model has bond dimension $\chi=5$. Compression can be achieved by either selecting a specific truncation parameter or by selecting a maximum bond dimension $\chi$. $H^2$ is constructed from multiplying two copies of the $H$ MPO. This is performed with the ITensors library using a direct MPO product to contract shared indices. Afterwards, for each power of $H$, one may sweep along the chain and truncate with a cutoff or a maximum bond dimension. As seen in Fig.~\ref{fig:HamiltonianCompression}, a first important result is that the bond dimensions are independent of system size. Furthermore, while the bare bond dimension of $H^K$ grows exponentially as $5^K$, for $10^{-6}$ approximation error the bond dimensions compressed operators are $10,16,24$ for $K=2,3,4$. The Hilbert-Schmidt inner product, implemented as a trace $\langle A, B \rangle\ = \rm Tr(A^\dagger B)$, is used to compute the Frobenius norm $||A||_F = \sqrt{\rm Tr(A^\dagger A)}$. The approximation error is then given by $||A-B||_F = \sqrt{\text{Tr}(A^\dagger A) + \text{Tr}(B^\dagger B)  - 2\text{Re}\{\text{Tr} (A^\dagger B) \}}$. Even large powers, such as $K=5,6,7$, are realizable with modest bond dimensions and modest relative Frobenius error $||H^K-H_{\chi}^K||_F/||H^K||_F$ between $10^{-4}$ and $10^{-2}$. To highlight the scalability of this method, note that, for $N=64$ and $K=7$, the MPO represents more than $8.61\times 10^{15}$ Pauli terms. 

\begin{figure}
    \centering
    \includegraphics[width=1\linewidth]{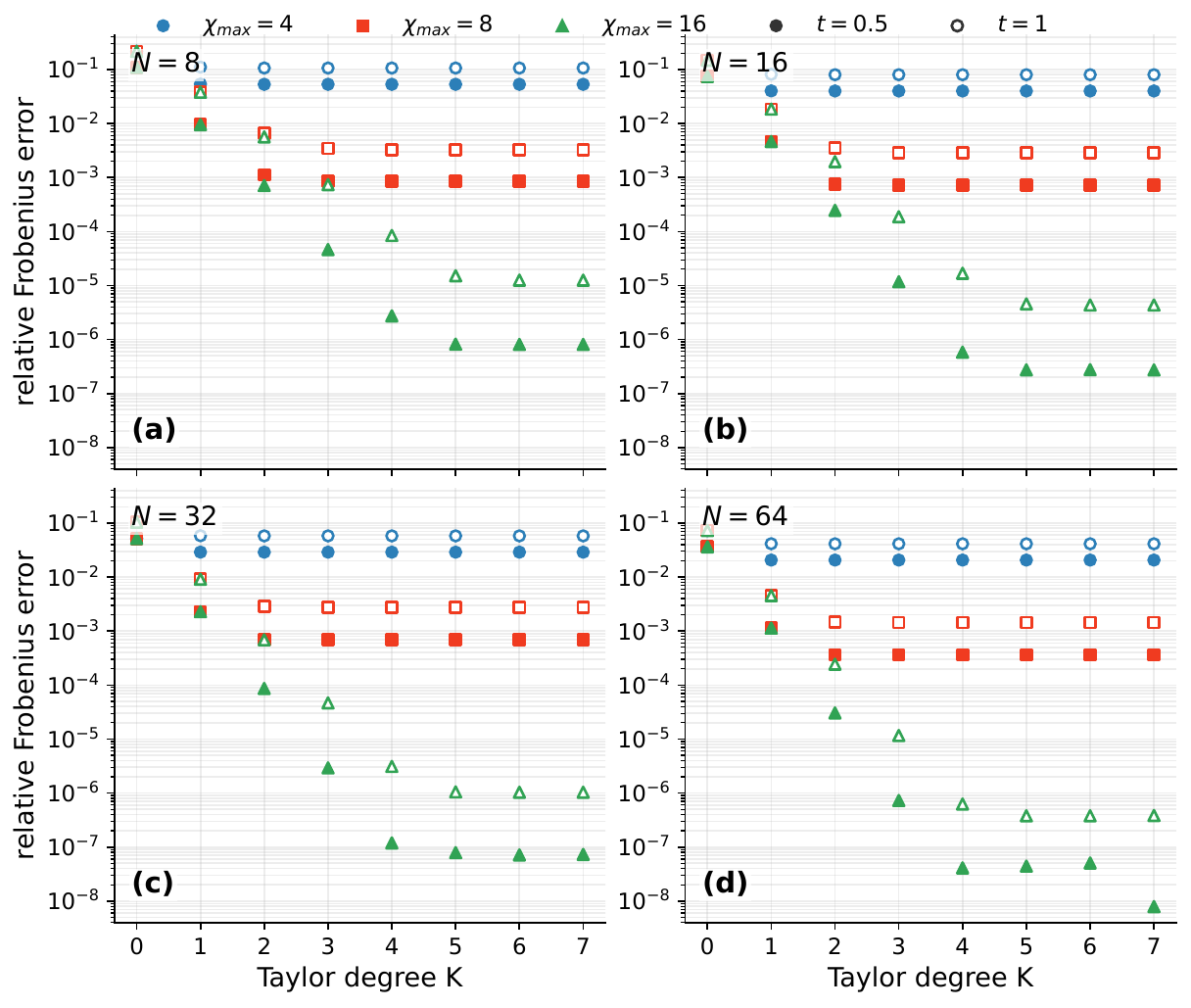}
    \caption{Relative Frobenius error in synthesizing Hamiltonian simulation unitaries as an MPO Taylor expansion, $||e^{-iHt} - \sum^K_{k=0} \frac{(-iHt)^k}{k!}||_F$, as a function of system size and for $t=0.5$ (open symbols) and $t=1$ (filled in symbols). The MPO bond dimension scaling is illustrated by the  circular, square, and triangular symbols which denote $\chi=4,8,16$ respectively.}
    \label{fig:MPOerrorscaling}
\end{figure}

For the real time evolution unitary synthesis, compression is even more effective. The reason is that higher orders are weighted by factors of $(-it)^k/k!$. As illustrated in Fig.~\ref{fig:MPOerrorscaling} this leads to even slower growth of the bond dimension. For system sizes of up to $N=64$ spins, a $\chi=16$ MPO sufficed in approximating the unitary up to a relative Frobenius error of $10^{-6}$. Note that the $N=8$ errors are computed with respect to a dense matrix exponentiation reference while for $N\geq16$ the errors are computed with respect to an MPO with bond dimension $128$ at order $K=8$. These results at larger system sizes highlight the MPOs rapid convergence in terms of bond dimension and, as expected for a Taylor series expansion~\cite{Childs2012}, in terms of $K$. 

\begin{figure}
    \centering
    \includegraphics[width=1\linewidth]{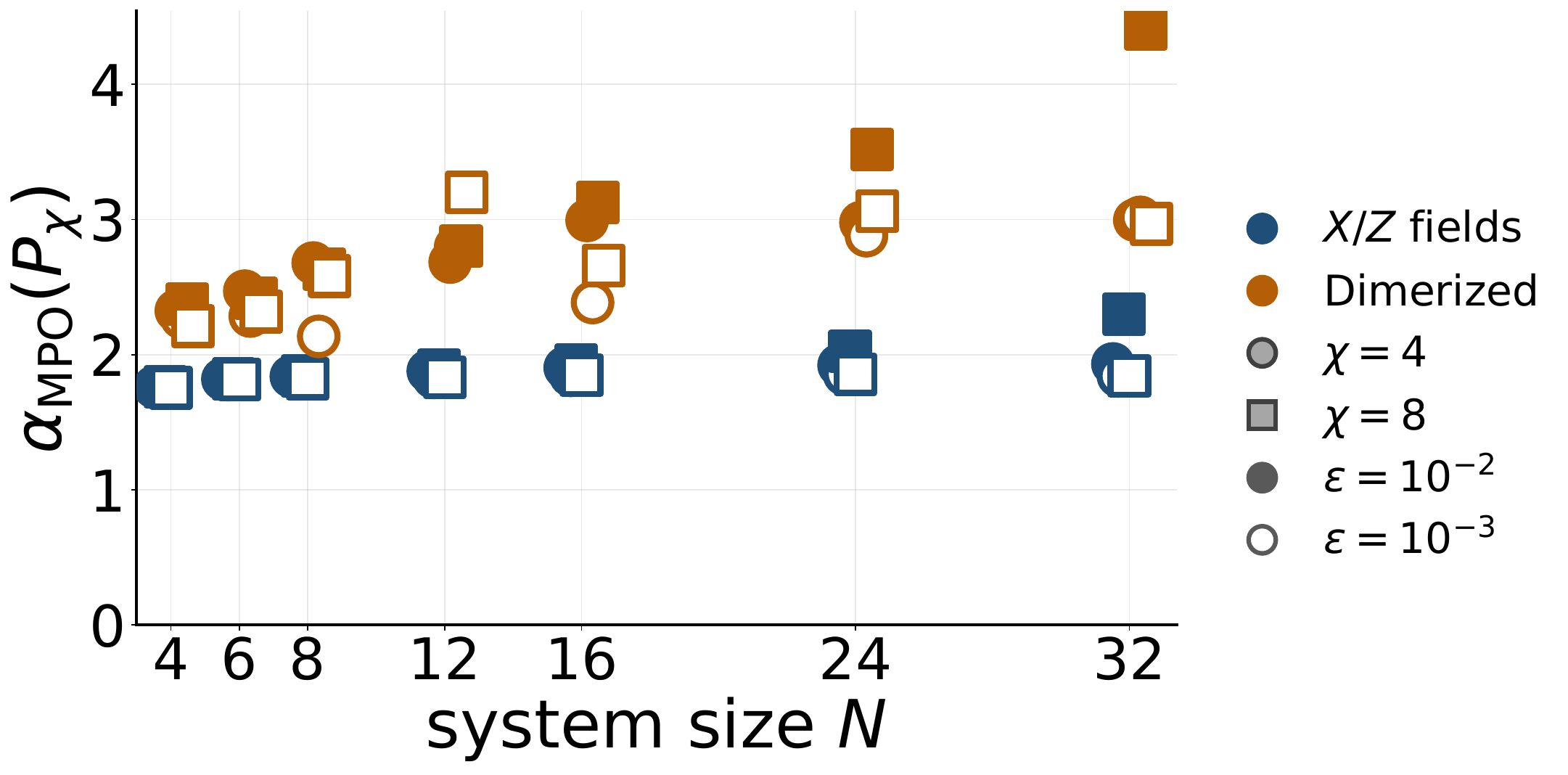}
    \caption{MPO block encoding normalization $\alpha_{\rm MPO}$. The blue (orange) markers represent the model with additional fields (dimerization) for $t=0.5$. Results for $\epsilon=10^{-2}$ ($10^{-3}$) are indicated by the filled in (open) markers. Circles and squares denote $\chi=4$ and $\chi=8$ respectively.}
    \label{fig:cost_scaling}
\end{figure}

To test whether the compiled-polynomial MPO route relies on the special uniform Heisenberg structure, we repeated the scan with two perturbed models. Namely, a Heisenberg chain with deterministic site-dependent \(X/Z\) fields, given by $H_2 = Eq.~\ref{eq:Heisenberg} + \sum_{i=1}^{N} \sum_{\mu=x,y,z} h_{\mu,i} S^\mu_{i}. $  For the magnetic fields, we consider an oscillating spin texture with deterministic site-dependent fields $h_{x,j} = 0.73 + 0.21 \sin(1.173j + 0.31), h_{y,j}=0, h_{z,j} = -0.47 + 0.19 \cos(0.719j + 0.83)$. Alternatively, we set $\delta=0.5$ and consider a dimerized Heisenberg chain with alternating bond strengths $J_i = J + \delta(-1)^i$. For each of these, we compute $\alpha_{\rm MPO}$ and the subsequent MPO-transition encoding cost $C_{\ref{eq:MPO_expansion}}$. As illustrated by Fig.~\ref{fig:cost_scaling}, for both families, the compiled approximately unitary MPO retained mild normalization. 

The cost of one MPO-transition LCU call is \(C_{\rm LCU}(P_\chi) = 2C_{\CPREP}+C_{\PHASE}+C_{\SELECT}\). For approximately unitary \(P_\chi(t)\), oblivious amplitude amplification gives the amplified application cost \( C_{\ref{eq:MPO_expansion}} \equiv C_{\rm apply}(P_\chi) \sim \alpha_{\rm MPO}(P_\chi(t))\,C_{\rm LCU}(P_\chi)\). To evaluate this proxy, we compile an MPO \(P_\chi(t)\), extract the transition coefficients \(W^{[n],\mu}_{a,b}\), compute the backward messages, and contract the resulting absolute-value transfer matrices to obtain \(\alpha_{\rm MPO}(P_\chi)\). These data define the factorized LCU circuit with conditional preparations, transition phases, and locally selected Pauli operators. The resulting cost is a logical oracle proxy at the \(\CPREP/\PHASE/\SELECT\) abstraction level, not a fault-tolerant gate count.

At \(t=0.5\) and \(\epsilon=10^{-3}\), the amplified MPO-transition application-cost proxy, \(C_{\ref{eq:MPO_expansion}}\equiv C_{\rm apply}(P_\chi) \sim \alpha_{\rm MPO}(P_\chi)C_{\rm LCU}(P_\chi)\), scaled as \(N^{1.17}\) for the field-perturbed Heisenberg model and \(N^{1.35}\) for the dimerized model at \(\chi=4\). At \(\chi=8\), the corresponding fitted exponents increased to \(N^{1.50}\) and \(N^{1.79}\), respectively. Thus the MPO-transition route remains close to linear at \(\chi=4\), while the larger-\(\chi\) data show stronger residual \(N\)-dependence, especially for the dimerized model. These experiments support the compiler-level claim that the favorable normalization of \(P_\chi(t)\approx e^{-itH}\) is not limited to the integrable uniform chain. 

To summarize, at fixed normalized time and fixed target accuracy, the scaling advantage of the MPO-transition route over the fully enumerated explicit Pauli Taylor LCU baseline is captured by
\begin{equation}
\frac{C_{\ref{eq:pauli_expansion}}} {C_{\ref{eq:MPO_expansion}}} =  \mathcal{O}\!\left(N^{K(t,\epsilon)-\nu}\right),
\end{equation}
where \(C_{\ref{eq:pauli_expansion}}=\mathcal{O}\!\left(N^{K(t,\epsilon)}\right)\) and \(C_{\ref{eq:MPO_expansion}}\sim N^\nu\), with \(\nu\in[1.17,1.79]\) across the models and bond-dimension budgets studied. This scaling holds when the MPO bond dimension and normalization remain mild.

\section{Conclusions}
\label{sec:conclusions}

In this work we introduced a new virtual-path LCU compiler for synthesizing and block encoding operators represented as matrix product operators. For the Hamiltonian-simulation examples considered here, the real-time propagator is first approximated as a Taylor series of weighted MPOs and compressed to a controlled bond dimension. The resulting, approximately unitary, MPO is then compiled directly into conditional \texttt{PREP}, phase, and local \texttt{SELECT} stages. Under the logical-transition cost model analyzed here, this construction replaces the $\mathcal{O}(N^K)$ branch table of an explicitly enumerated Pauli-polynomial LCU by a quadratic polynomial in the MPO bond dimension. Our asymptotic comparison was performed against explicit Pauli-branch enumeration, not against all structured Taylor-LCU algorithms such as BCCKS~\cite{BCCKS2017} that avoided the cost through an exact Cartesian-product factorization. In future works, it will be interesting to determine whether our method can improve the compilation, in terms of gate-count,of block encoding tasks lacking such factorization properties.

This distinction suggests a set of broader next questions. That is, which tensor factorizations admit efficient \texttt{PREP} and \texttt{SELECT} constructions? The dense $\mathcal{O}(q\chi^2)$ conditional-preparation cost used here is an early result rather than a final limit. Further improvements may follow from sparsity, symmetry, local low-rank decompositions, canonical gauges, finite-state automaton structure, or hybrid factorizations that combine the history-space structure exploited by BCCKS with the spatial compression captured by MPOs. Determining when these structures reduce quantum gate count is an important direction for future work.

Full end-to-end comparisons with qubitization algorithms remains an important direction. Quantum eigenvalue transformation pays the time-and-precision dependence through the polynomial degree $D(\alpha t,\epsilon)$ and repeated calls to a block encoding of \(H\). The compiled-MPO route instead pays this dependence through the Taylor or polynomial order \(K(t,\epsilon)\), the bond dimension \(\chi_P(K,t,\epsilon)\), and the MPO-transition normalization \(\alpha_{\rm MPO}(P_{\chi,K}(t))\) of the compressed finite-time propagator. A fair comparison must therefore combine the cost of the initial Hamiltonian oracle, the QET iterate count, the cost of conditional MPO preparation and selection, the growth of \(\chi_P\) with \(t\) and \(\epsilon\), and the resulting success-probability or amplitude-amplification overhead. It remains to be seen if, whether by using our virtual path LCU encoding or by other means, a suitable encoding is found in which the base operator norm is mildly increasing so that this encoding can be used as a subroutine for qubitization methods. We leave this holistic analysis for future work.

By generalizing our unitary synthesis from a time-independent polynomial in $H$ to the time-dependent case~\cite{vanthilt2026, VanDamme2026}, it will be interesting to compare the algorithmic scalings of block encoding time-dependent tensor network encodings to conventional methods which, for example, break up the unitary in terms of a series of time dependent unitaries. Another related direction we have not explored in this work is to first synthesize unitary circuits, which could be time independent or dependent, within a tensor network formalism without the use of Taylor expansions~\cite{Gibbs2025deepcircuit, Zhang2026}. This may open up new routes to utilize product formulas to first synthesize unitaries, which can, through the techniques we have described, then be block encoded which enables functions of them to be applied. 

Another direction is to combine the present MPO-transition compiler with multiproduct formulas which approximate real-time evolution by linearly combining product-formula circuits rather than by expanding the target unitary into Pauli strings~\cite{Childs2012,Low2019MPF}. Recent work has developed randomized~\cite{Faehrmann2022}, hardware-friendly hybrid~\cite{CarreraVazquez2023}, dynamic~\cite{Zhuk2024}, and tensor-network-enhanced~\cite{Robertson2025} variants of this idea. From the perspective of the present work, these methods expose a fourth LCU structure: the branches are entire product-formula circuits. It would be natural to ask whether tensor-network compression can be used either to select, compress, or precondition the product-formula branches, or conversely whether multiproduct formulas can provide a shallower circuit backend for the compiled MPO polynomial \(P_\chi(t)\). This comparison is distinct from the costs analyzed here and should depend on the conditioning of the multiproduct coefficients, the depth of the product-formula branches, and the MPO bond dimension and path normalization of \(P_\chi(t)\).

Finally, one could also apply our methods to non-unitary functions. An example of this was already provided by Ref.~\citenum{Siegl2026}, but this could be extended in a variety of ways. Other examples include the potential use to simulate time evolve in superposition across time-scales~\cite{Choi2021} similarly to quantum phase estimation or to implement the Hubbard-Stratonovic~\cite{Keen2021, Bell2025} integration of the real time evolution. Lastly, this work focused on encodings of approximately unitary real-time propagators. Since these examples generally leverage non-unitary operator synthesis, a holistic resource analysis including success probability scaling would be required to evaluate which approach is suitable to each use case. 

\section{Acknowledgments}

E.D. acknowledges constructive conversations with Y. Wang. E.D. is supported by the U.S. Department of Energy, Office of Science, Advanced Scientific Research Program, Early Career Award under contract number ERKJ420.

\bibliographystyle{unsrt}
\bibliography{refs}

\end{document}